# Evolution of the Ubiquitin-Activating Enzyme Uba1 (E1)

Douglas C. Allan[1] and J. C. Phillips[2]

1. Corning Inc, Div. Sci. & Technol., Corning, NY 14831 USA

2. Dept. of Physics and Astronomy, Rutgers Univ., Piscataway, N. J., 08854

Abstract

Ubiquitin tags diseased proteins and initiates an enzyme conjugation cascade, which has three stages. The first-stage enzyme Uba1 (E1) has evolved only modestly from slime mold to humans, and is > 14 times larger than Ub. Here we use critical point thermodynamic scaling theory to connect Uba1 (E1) evolution from yeast and slime mold to fruit flies and humans to subtle changes in its amino acid sequences.

**Introduction**

The ubiquitin-activating enzyme Uba1 (E1) constitutes the first step in the covalent cascade modification of target proteins with ubiquitin (Ub). Ubiquitin itself, discovered less than 50 years ago, tags thousands of diseased proteins for destruction [1,2]. It is small (only 76 amino acids), and is found unchanged in mammals, birds, fish and even worms. Because of its universality, Ub is a valuable proving ground for universal biophysical theories discussing protein amino acid sequences, structure and function [3]. Indeed key features of Ub functionality (hydropathic waves) were identified using critical point thermodynamic scaling theory [4]. The general biochemical logistics of Ub activation, conjugation and ligation are orchestrated sequentially by the Ub conjugation cascade of E1, E2 and E3 enzymes. Humans are known to harbor two E1, ~30 E2 and ~600 E3 enzymes in the Ub conjugation cascade [5]. While Ub is "perfect", Uba1 (E1) has evolved only modestly from slime mold to humans. The details of this evolution express several leading features of enzyme functionality. Uba1 (E1) is a large protein (> 1000 amino acids), but it is readily treated by critical point thermodynamic scaling theory, with its firm foundations in statistical mechanics and its bioinformatically determined universal parameters [3]. It turns out that hydropathic waves are also useful for Uba1 (E1), which is > 14 times larger than Ub.

As before [3,4], all calculations are based on thermodynamically first- and second-order hydropathic $\Psi$(amino acid) scales [6,7], linearly scaled to a common center and a common range for each of the 20 amino acids. These are then converted to a triangular matrix $\Psi$(aa,W), where W is the length of a sliding window centered on each amino acid site. We have studied the range $1 \leq W \leq 75$, which is includes values of W much larger than the small value, fixed at W ~ 5, in most calculations using sliding windows [8]. Just as one focuses a microscope to optimize its image, one scans W to optimize its recognition of allometric regularities of hydropathic hot spots (hydrophobic extrema of $\Psi$(aa,W)) at special values of W = W*.

**Results**

The hydrophobic extrema of neuroglobin form sophisticated patterns that are closely related to the evolution of specific species. For example, mouse and rabbit escape predators in different ways, and these differences are recognizable in their $\Psi$(aa,W*) profiles [9]. There are several other examples already of proteins whose hydrophobic extrema form level sets. In Fig. 1 we plot the profiles of Uba1 (E1) for humans and slime mold. The choice W* = 55 levels the human hydrophobic extrema, and simultaneously aligns the slime mold extrema linearly with a small tilt (about 15% of the overall range). Such excellent alignments (to within 1%) are unlikely and not accidental. For instance, the differences between the MZ and KD scales are small (85% correlation [10]), yet as Fig. 2 shows, the successful pivotal alignment with the MZ scale is lost with the KD scale.

Structural data are most complete for Uba1 (E1) yeast, and the human and yeast profiles are compared in Fig. 3. The differences are small, and are mentioned in the Fig. 3 caption. Before we compare the long-range ("allosteric") correlations of these figures, we show in Fig. 4 the results for fruit fly, which has a lifetime of days, not years. This implies that its Uba1 kinetics are ~ $10^3$ faster than human Uba1 kinetics. It is plausible that the two hydrophilic minima discussed in Fig. 4, which are 3-6 lower values of Ψ(aa,W*) (5-10% of the full range) than in the human Ψ(aa,W*), are good indicators of this kinetics acceleration.

**Discussion**

Before comparing our results with structural studies, we can turn to the Wiki on transition state theory (1935), which explains the reaction rates of elementary chemical reactions in terms of two parameters in one dimension. Structural studies contain information on the ground state, a minimum in configuration space, whereas rates are determined by the properties of transition states, technically also saddle points in configuration space. Both minima and saddle points are also thermodynamic critical points, where long-range attractive and short-range repulsive interactions are equal at the critical temperature, close to body temperature. Our conjecture here is that studying Ψ(aa,W*) extrema, both the hydrophobic pivots and the hydrophilic hinges [3], provides one-dimensional insights into extremal sets of two parameters describing transition state properties, which can be compared to structural studies [11,12] of ground state properties. Because Uba1 is so large, the free energy differences between these two states are very small, and the static and dynamic structural differences are expected to involve interdomain conformational motion.

According to [11], Uba1-E1 consists of four building blocks: first, the adenylation domains composed of two motifs (labelled IAD (1-169) and AAD (404-594), for "inactive" and "active" adenylation domain, respectively), the latter of which binds ATP and Ub; second, the catalytic cysteine half-domains, which contain the E1 active site cysteine (CC (169-268) and CCD (594-860) inserted into each of the adenylation domains; third, a four-helix bundle 4HB (268-356) that represents a second insertion in the IAD; and fourth, the C-terminal ubiquitin-fold domain (UFD (926-1024)), which recruits specific E2s. How do these structural and functional domains compare with our one-dimensional hydropathic profiles? Fig. 5 shows an excellent match, with the domain boundaries associated either with pivots or one amphiphilic side of a hinge. One could argue here that the separation of 4HB is unnecessary, but this secondary structure includes the deep and wide hydrophilic minimum 330-370, with a sharp edge at 320 of Ψ(aa,W*).

The differences between ground and transition state properties are exemplified by a comparison of our human-yeast results with those of a structural study of possible differences in ubiquitin-Uba-1 binding in yeast (known structure) with human (simulated in [12]). Their discussion focuses on Ubiquitin contacts with Tyr 571 (618 in our site numbering) in the CCD domain. According to Fig. 3 and its caption, there are substantial differences between yeast and human profiles here, and they are related to differences in tilts of the hydrophobic pivots. These are consistent with the modelling results, and also show the importance of allometric interactions between folded CC and CCD domains (Fig. 5).

Human AAD adenylation (ATP) sites are 478, 504, 515, and 528. The overall BLAST identities and positives for yeast and human Uba1 are 52% and 71%, and these increase in the 400-600 ADD range to 63% and 80%. This range is shown in Fig. 6, and we see strong similarities in the two $\Psi(aa,W^*)$ profiles. Their correlation is 86%, which means that $\Psi(aa,W^*)$ is more effective in the 400-600 ADD range than even BLAST positives (80%) because it uses the MZ scale and because $W^*$ has been chosen to display level set hydrophobic adenylation domain pivots. From Fig. 6 we see that the four sites are concentrated in the center of the ADD structural domain. They line the amphiphilic range from the hydrophobic pivot at 478 to close to the hydrophilic hinge at 535. The profile $\Psi(aa, 55)$ not only displays a stronger yeast-human ADD correlation than BLAST, but it also has revealed a hydropathic cascade of ATP binding sites. Among mammals the similarities are of course greater. The overall correlation of the mouse and human $\Psi(aa,W^*)$ profiles is 97.2%, which increases in the 400-600 ADD range to 98.6 %.

There has recently been some interest in Uba6, which is most similar to slime mold, with BLAST identities of 59% and positives 73% in the 400-600 ADD range. The 400-600 ADD correlation of the two $\Psi(aa,W^*)$ profiles is a striking 87%, so functional differences probably arise outside the ADD binding domain. Human Uba6 and Ube1 have distinct preferences for E2 charging in vitro, and their specificity depends in part on their C-terminal ubiquitin-fold domains, which recruit E2s [13]. Comparison of Uba1 of yeast and slime mold with Uba6 shows that for the most part, where Uba6 differs from one of the two, it is similar to the other. The exceptional region is UFD, the ubiquitin fold domain (Fig. 5), where Uba6 has a strong hydrophobic peak.  Similar differences are obvious in the human and fruit fly profiles.  This "only" confirms the main conclusion of [13], but note that it does so through a simple one-dimensional analysis that also includes many species' versions of Uba1.

**Conclusions**

At present the most sophisticated structural studies involving MDS simulations with explicit water [13] can reveal short-range interspecies differences in Uba-E1 – Ub binding, but long-range interactions are not identified.  Proteins are near thermodynamic critical points in amino acid configuration space, and near such points (especially transition states),  long-range and short-range interactions are balanced.  In many proteins the short-range interactions evolve in subtle ways inaccessible to experiment, while here we have shown that the long-range interactions change in ways that can be easily recognized in $\Psi(aa, W^*)$.

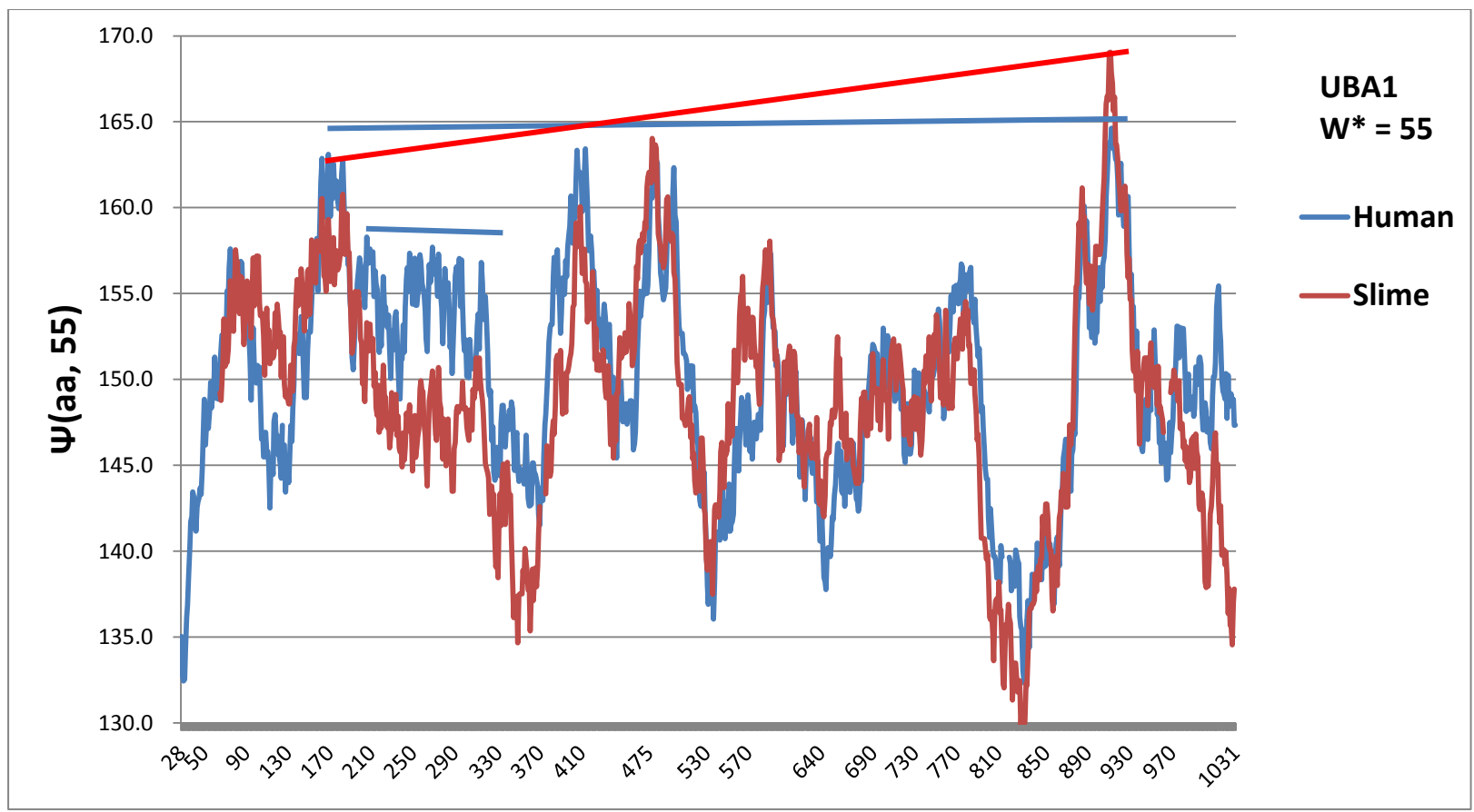

Fig. 1. The $\Psi(aa,W*)$ profiles for human and slime mold Uba1 (E1), using the modern MZ scale [7], which describes second-order (allosteric) evolutionary approaches to perfection (thermodynamic critical point [3]). A similarly successful value for W* using the standard KD classical (thermodynamically first-order) scale [6] was not found. Note the secondary leveling of the human profile between sites 200 and 330. The site numbering here is that of Uniprot P22314 (UBA_1HUMAN).

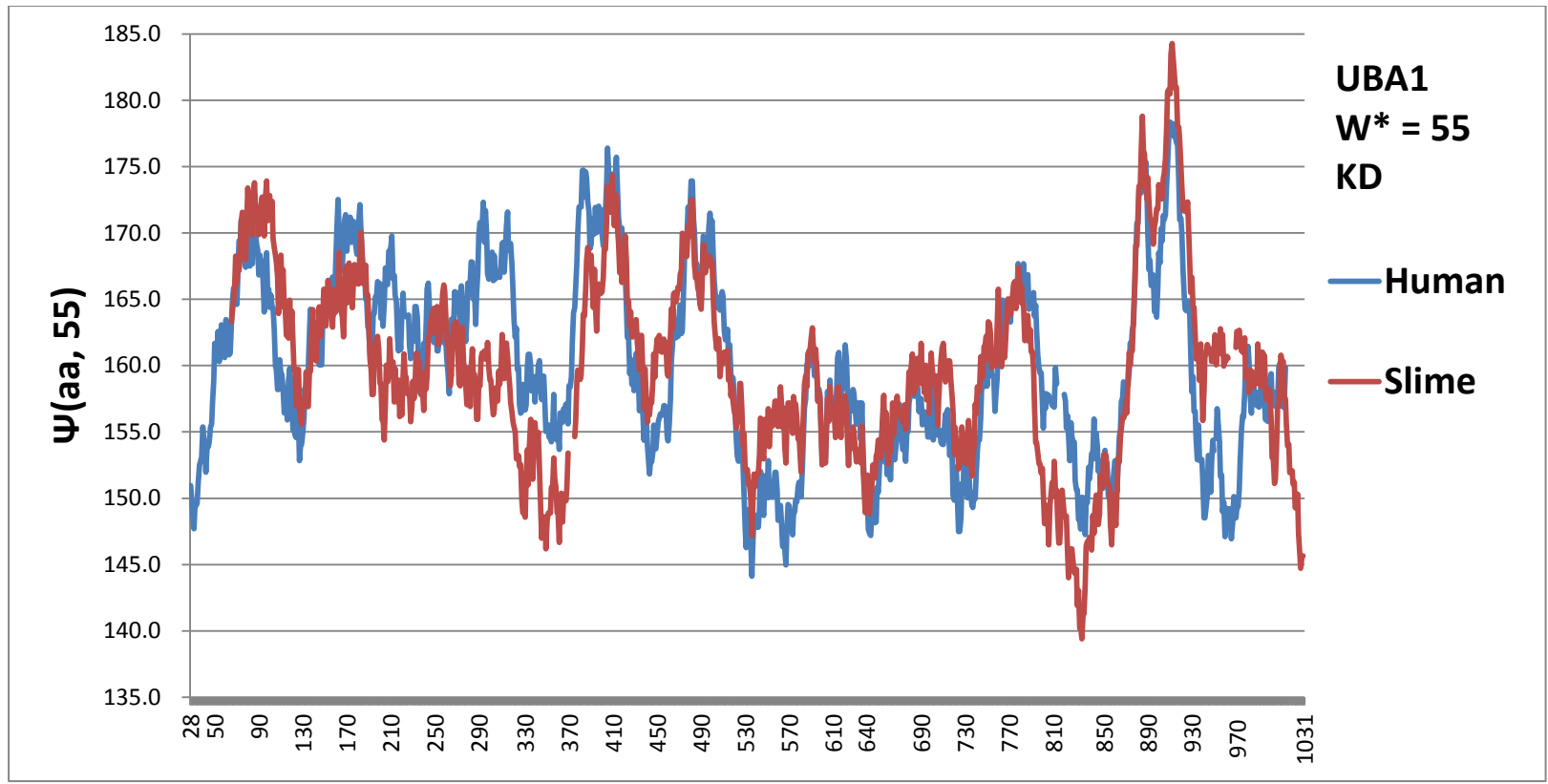

Fig. 2. The alignment of the hydrophobic extrema found with the MZ scale in Fig. 1 does not occur with the KD scale. The two sequences were aligned using BLAST, which found a few short gaps. One of these is seen here in the slime mold curve near 370, and two others can be seen in human near 810 and slime near 950.

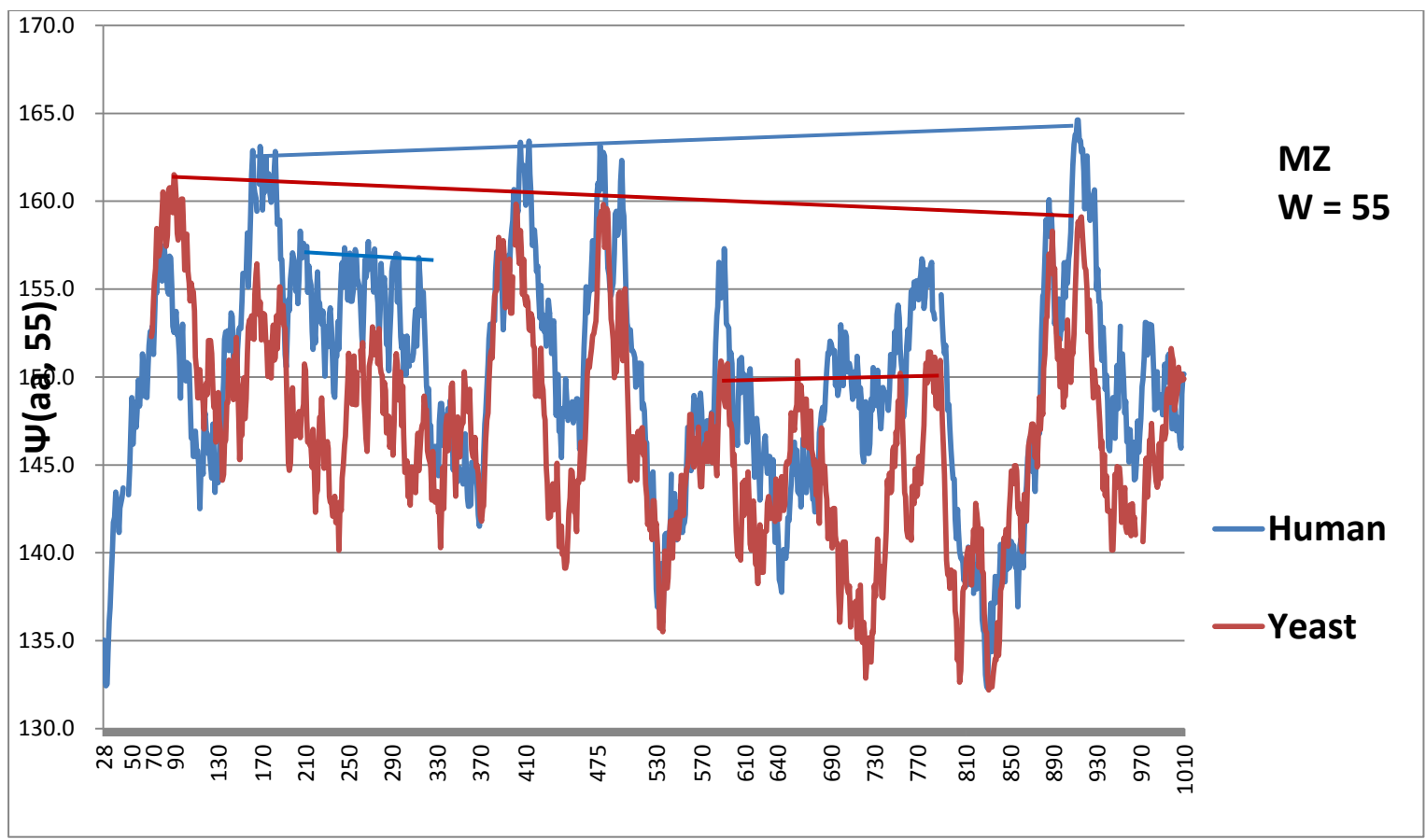

Fig. 3. Yeast and human UBA1hydrophobic pivots are closely similar, and both have only small amphiphilic pivotal tilts with opposite signs. The secondary leveling seen in the human profile between sites 200 and 330 is absent in yeast, which contains a different secondary leveling between 600 and 800. The reversal of the two secondary alignments is consistent with the opposite signs of the pivotal tilts. In each case the secondary leveling is associated with the slightly more hydrophilic protein half. Yeast has several unusually deep hydrophilic hinges near 220, 460 and 690.

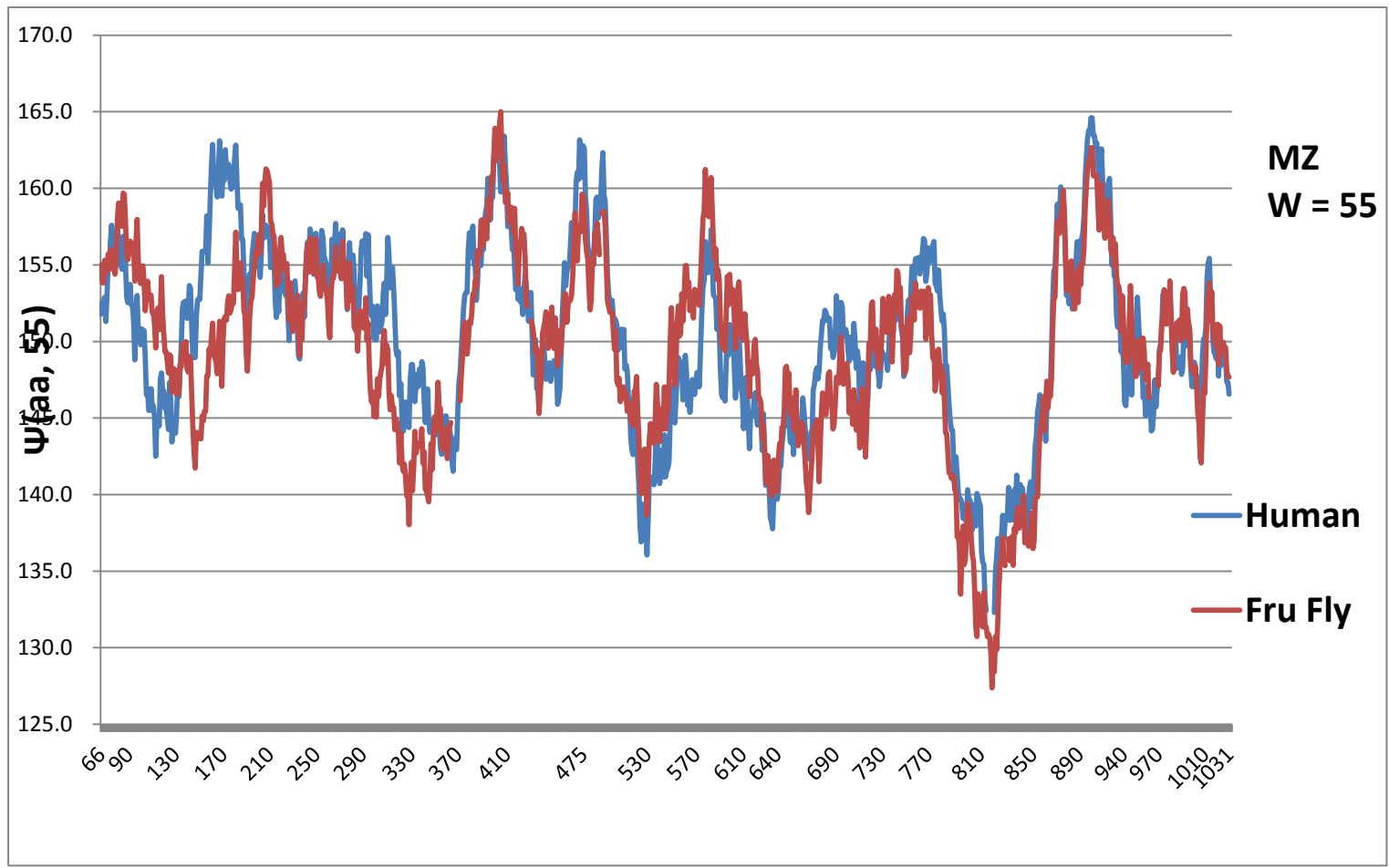

Fig. 4. Aligned profiles of Uba1 human and fruit fly are similar, but with significant differences. Note especially the deeper hydrophilic minima of fruit fly near 815, where there is a six amino acid gap in the aligned human profile, and at 315. There is a six amino acid gap in the aligned fruit fly profile at 370, close to the human minimum, which presumably is associated with the deeper fruit fly minimum.

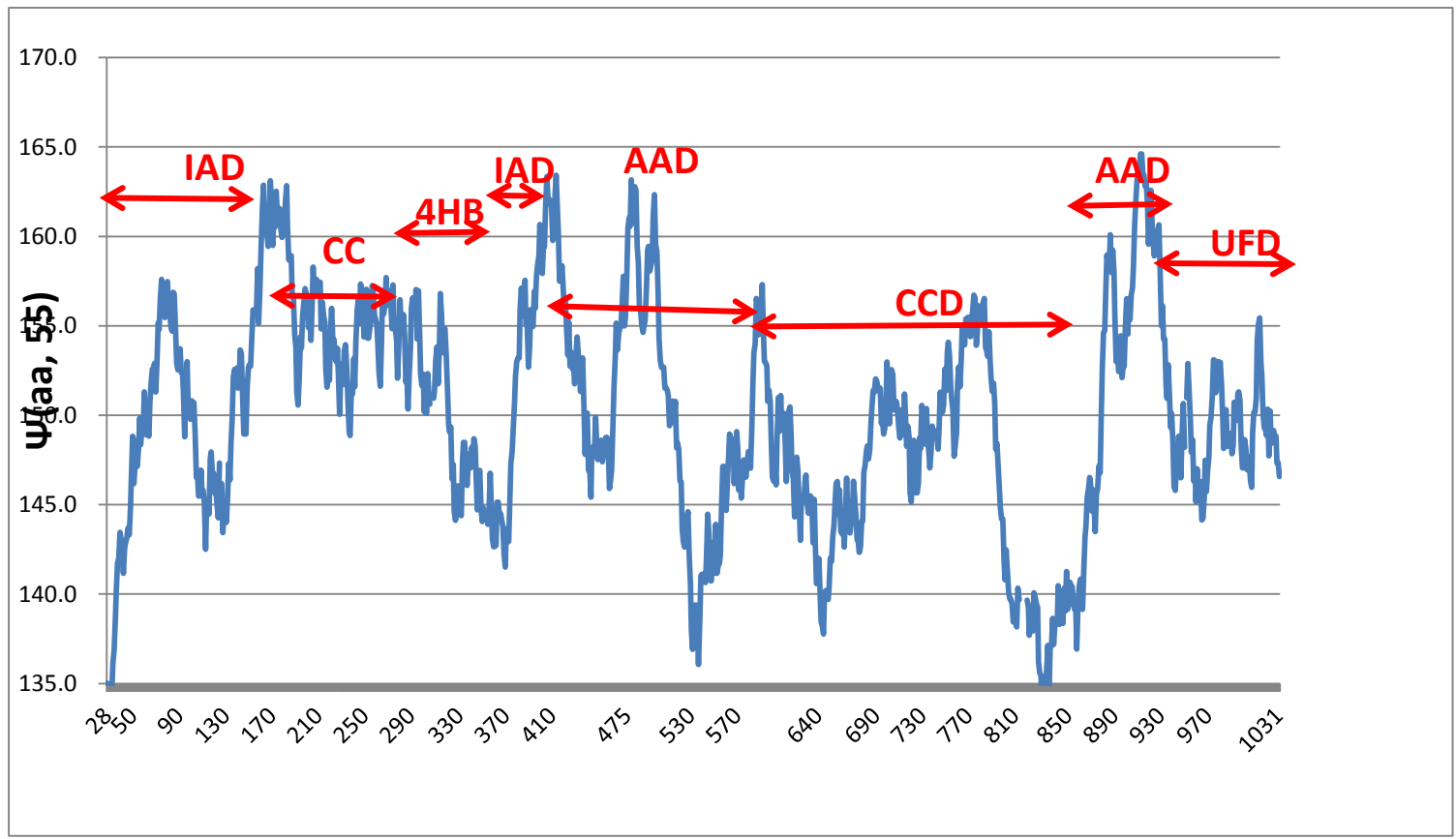

Fig. 5. Labelling of the human MZ profile Ψ(aa, 55) according to the structural analysis of [11] reveals many very accurate pivotal set leveling effects. The most obvious one is the four highest **pivoting** peaks (Fig. 1), associated with the two parts of IAD and AAD, ''inactive'' and ''active'' adenylation domains. Similar leveling occurs in yeast (Fig. 3), and somewhat more weakly in fruit fly (Fig 4), while the **adenylation pivoting** set in slime mold is tilted but still linear (Fig. 1).

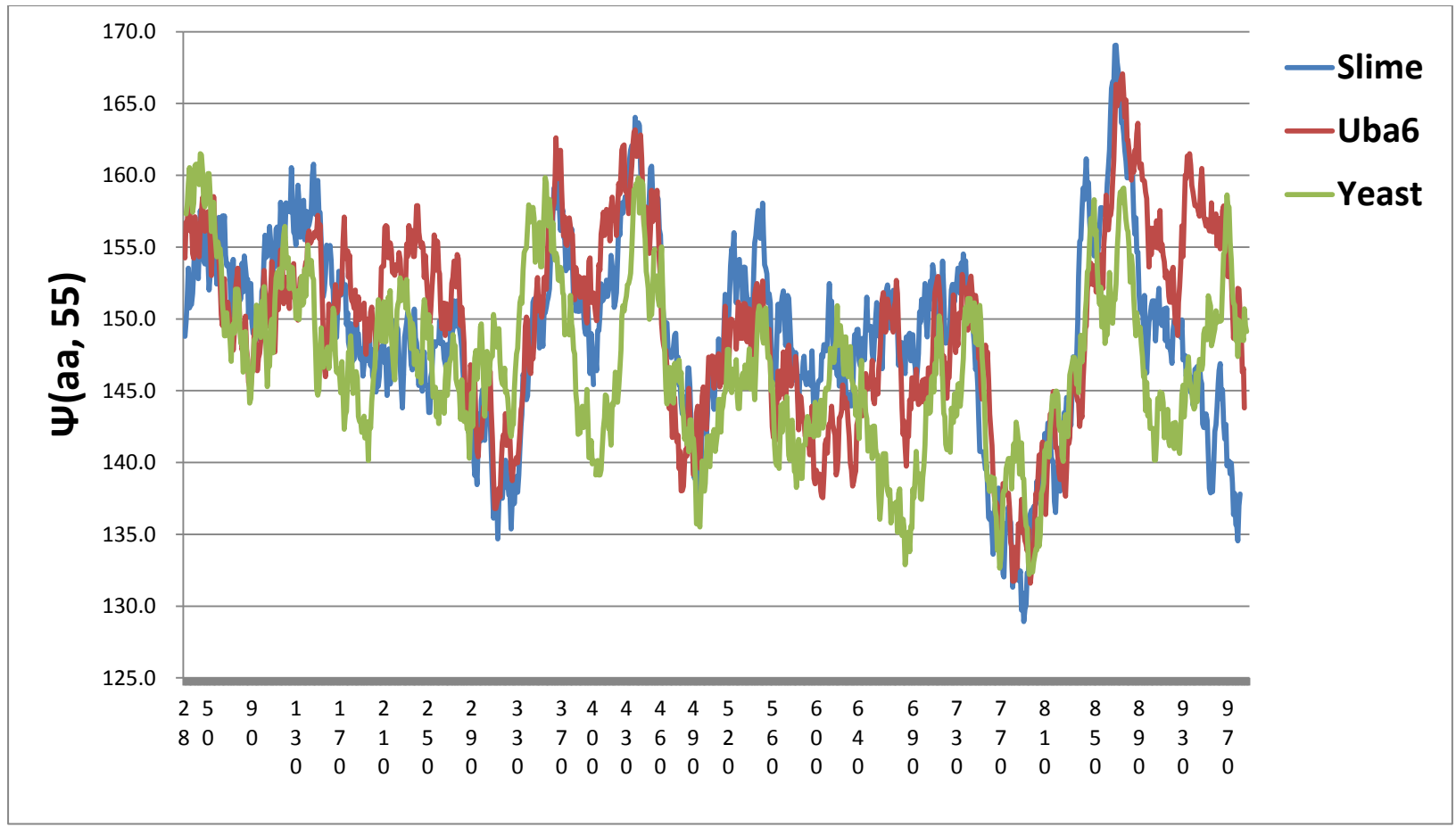

Fig. 6. Comparison of Uba1 of yeast and slime mold with Uba6 shows that for the most part, where Uba6 differs from one of the two, it is similar to the other. The exceptional region is UFD, the ubiquitin fold domain (Fig. 5), where Uba6 has a strong hydrophobic peak near 942.